\documentclass[conference, letterpaper, 10pt]{IEEEtran}
\usepackage{amsmath,amssymb,amsfonts}
\usepackage{graphicx}
\usepackage{textcomp}
\usepackage{xcolor}
\def\BibTeX{{\rm B\kern-.05em{\sc i\kern-.025em b}\kern-.08em
    T\kern-.1667em\lower.7ex\hbox{E}\kern-.125emX}}

\usepackage{mathtools}
\usepackage[normalem]{ulem}
\usepackage[frozencache=true,cachedir=minted-cache]{minted}
\usepackage{tcolorbox}
\usepackage[bookmarks=false]{hyperref}
\usepackage[hyphenbreaks]{breakurl}
\usepackage{siunitx}
\usepackage{listings}
\usepackage{multirow}
\usepackage{pgfplots}
\usepackage{pgfplotstable}
\pgfplotsset{compat=1.18, empty line=none}
\DeclareSIUnit{\sample}{Sa}
\DeclareSIUnit{\bit}{b}
\DeclareSIUnit{\byte}{B}
\DeclareSIUnit{\cycle}{cycle}
\DeclareSIUnit{\cycles}{cycles}
\DeclareSIUnit{\one}{1}
\usepackage[inline]{enumitem}
\usepackage{algorithm}
\usepackage{algpseudocode}
\algblockdefx[ForLab]{ForLab}{EndForLab}%
    [2]{\emph{#1:} \textbf{for} #2 \textbf{do}}%
    {\textbf{end for}}
\algnewcommand{\LeftComment}[1]{\Statex \(\triangleright\) #1}
\usepackage{inconsolata}
\usepackage{booktabs}
\usepackage[capitalize]{cleveref}
\crefformat{equation}{(#2#1#3)}
\crefmultiformat{equation}{(#2#1#3)}{ and~(#2#1#3)}{, (#2#1#3)}{ and~(#2#1#3)}
\crefrangemultiformat{equation}{(#3#1#4)--(#5#2#6)}{,(#3#1#4)--(#5#2#6)}{,(#3#1#4)--(#5#2#6)}{,(#3#1#4)--(#5#2#6)}
\crefrangeformat{equation}{(#3#1#4)--(#5#2#6)}
\usepackage[style=ieee,minbibnames=100,maxbibnames=100,mincitenames=1,maxcitenames=1]{biblatex}
\usepackage{tikz}
\usepackage{microtype}
\usepackage{standalone}
\ifCLASSOPTIONcompsoc
\usepackage[caption=false,font=normalsize,labelfont=sf,textfont=sf]{subfig}
\else
\usepackage[caption=false,font=footnotesize]{subfig}
\fi
\usepackage{mycolors}
\usepackage{tikz}
\usepackage{tikz-timing}
\usetikzlibrary{positioning}
\usetikzlibrary{shapes}
\usetikzlibrary{fit}
\usetikzlibrary{arrows.meta}
\usepackage{acronym}

\makeatletter
\newcommand*{\org@overidelabel}{}
\let\org@overridelabel\@verridelabel
\@ifpackagelater{acronym}{2015/03/21}{
  \renewcommand*{\@verridelabel}[1]{%
    \@bsphack
    \protected@write\@auxout{}{\string\AC@undonewlabel{#1@cref}}%
    \org@overridelabel{#1}%
    \@esphack
  }%
}{
  \renewcommand*{\@verridelabel}[1]{%
    \@bsphack
    \protected@write\@auxout{}{\string\undonewlabel{#1@cref}}%
    \org@overridelabel{#1}%
    \@esphack
  }%
}
\makeatother

\acrodef{API}{application programming interface}
\acrodef{ASIC}{application-specific integrated circuit}
\acrodefindefinite{ASIC}{an}{an}
\acrodef{AXI}{advanced extensible interface}
\acrodef{BLAS}{basic linear algebra}
\acrodef{BRAM}{block random access memory}
\acrodefplural{BRAM}{block random access memories}
\acrodef{BW}{bandwidth}
\acrodef{CDC}{clock domain crossing}
\acrodef{CDFG}{control/dataflow graph}
\acrodef{DDG}{data dependence graph}
\acrodef{DDR}{double-data-rate}
\acrodef{DRAM}{dynamic random access memory}
\acrodefplural{DRAM}{dynamic random access memories}
\acrodef{DSP}{digital signal processor}
\acrodef{DFG}{dataflow graph}
\acrodef{DSE}{design-space exploration}
\acrodef{DUT}{design under test}
\acrodefplural{DUT}{designs under test}
\acrodef{EDA}{electronic design automation}
\acrodef{FF}{flip-flop}
\acrodef{FIFO}{first-in-first-out}
\acrodef{FPGA}{field-programmable gate array}
\acrodefindefinite{FPGA}{an}{an}
\acrodef{FPS}{frames per second}
\acrodefindefinite{FPS}{an}{an}
\acrodef{FU}{functional unit}
\acrodefindefinite{FU}{an}{an}
\acrodef{GEMV}{general matrix vector}
\acrodef{HLS}{high-level synthesis}
\acrodef{HW}{hardware}
\acrodef{II}{initiation interval}
\acrodefindefinite{II}{an}{an}
\acrodef{IP}{intellectual property}
\acrodefindefinite{IP}{an}{an}
\acrodef{IPI}{intellectual property integrator}
\acrodefindefinite{IPI}{an}{an}
\acrodef{LCS}{load-compute-store}
\acrodefindefinite{LCS}{an}{an}
\acrodef{LUT}{look-up table}
\acrodefindefinite{LUT}{an}{an}
\acrodef{MAC}{multiply and accumulate}
\acrodef{MIPS}{mega-instructions per second}
\acrodef{MCDFG}{multi-clock dataflow graph}
\acrodefindefinite{MCDFG}{an}{a}
\acrodef{MUL}{multiplier}
\acrodef{OP}{operation}
\acrodefindefinite{OP}{an}{an}
\acrodef{PIPO}{ping-pong}
\acrodef{PPA}{power, performance, and area}
\acrodef{QoR}{quality of results}
\acrodef{RTL}{register-transfer level}
\acrodefindefinite{RTL}{an}{an}
\acrodef{SDC}{system of difference constraints}
\acrodefindefinite{SDC}{an}{an}
\acrodef{SDFG}{synchronous dataflow graph}
\acrodefindefinite{SDFG}{an}{an}
\acrodef{SoC}{system-on-chip}
\acrodefindefinite{SoC}{an}{an}
\acrodef{SOTA}{state-of-the-art}
\acrodefindefinite{SOTA}{an}{a}
\acrodef{SCDFG}{single-clock dataflow graph}
\acrodefindefinite{SCDFG}{an}{a}
\acrodef{SRAM}{static random access memory}
\acrodefplural{SRAM}{static random access memories}
\acrodef{SW}{software}
\acrodef{VMS}{virtual molecule screening}

\usepackage[draft,inline,nomargin,noindex]{fixme}
\fxsetup{theme=color,mode=multiuser}
\FXRegisterAuthor{mtl}{amtl}{\color{blue}MTL}
\FXRegisterAuthor{gb}{agb}{\color{myorange}GB}
\FXRegisterAuthor{ll}{all}{\color{green}[LL}
\FXRegisterAuthor{fo}{afo}{\color{violet}[FO}

\makeatletter
\pgfdeclareshape{document}{
\inheritsavedanchors[from=rectangle] 
\inheritanchorborder[from=rectangle]
\inheritanchor[from=rectangle]{center}
\inheritanchor[from=rectangle]{north}
\inheritanchor[from=rectangle]{south}
\inheritanchor[from=rectangle]{west}
\inheritanchor[from=rectangle]{east}
\backgroundpath{
\southwest \pgf@xa=\pgf@x \pgf@ya=\pgf@y
\northeast \pgf@xb=\pgf@x \pgf@yb=\pgf@y
\pgf@xc=\pgf@xb \advance\pgf@xc by-10pt 
\pgf@yc=\pgf@yb \advance\pgf@yc by-10pt
\pgfpathmoveto{\pgfpoint{\pgf@xa}{\pgf@ya}}
\pgfpathlineto{\pgfpoint{\pgf@xa}{\pgf@yb}}
\pgfpathlineto{\pgfpoint{\pgf@xc}{\pgf@yb}}
\pgfpathlineto{\pgfpoint{\pgf@xb}{\pgf@yc}}
\pgfpathlineto{\pgfpoint{\pgf@xb}{\pgf@ya}}
\pgfpathclose
\pgfpathmoveto{\pgfpoint{\pgf@xc}{\pgf@yb}}
\pgfpathlineto{\pgfpoint{\pgf@xc}{\pgf@yc}}
\pgfpathlineto{\pgfpoint{\pgf@xb}{\pgf@yc}}
\pgfpathlineto{\pgfpoint{\pgf@xc}{\pgf@yc}}
}
}
\makeatother

\newcommand{\nodes}{
\begin{tikzpicture}
\node[draw, fill=mygrey!20, minimum height=.6cm, minimum width=.6cm] (v_0) {$v_0$};
\node[draw, fill=mygrey!20, xshift=-.1cm, yshift=-.1cm, minimum height=.6cm, minimum width=.6cm] at (v_0) (v_1) {$v_1$};
\node[draw, fill=mygrey!20, xshift=-.1cm, yshift=-.1cm, minimum height=.6cm, minimum width=.6cm] at (v_1) (v_i) {$v_i$};
\end{tikzpicture}
}

\pgfdeclarelayer{background}
\pgfdeclarelayer{floating}
\pgfsetlayers{background,main,floating}

\colorlet{read_x_color}{mygreen!30}
\colorlet{mult_0_color}{myorange!30}
\colorlet{mult_1_color}{myblue!30}
\colorlet{write_y_color}{mybrown!30}

\colorlet{major_mod}{black}
\colorlet{maybe_delete}{black}

\usepackage{ifthen}
\makeatletter
\newcounter{IEEE@bibentries}
\renewcommand\IEEEtriggeratref[1]{%
  \renewbibmacro{finentry}{%
    \stepcounter{IEEE@bibentries}%
    \ifthenelse{\equal{\value{IEEE@bibentries}}{#1}}
    {\finentry\@IEEEtriggercmd}
    {\finentry}%
  }%
}
\makeatother

\begin{document}
\title{A DSP shared is a DSP earned:\\
HLS Task-Level Multi-Pumping for\\
High-Performance Low-Resource Designs}

\author{
\IEEEauthorblockN{Giovanni Brignone, Mihai T. Lazarescu, Luciano Lavagno}
\IEEEauthorblockA{\textit{Dipartimento di Elettronica e Telecomunicazioni} \\
\textit{Politecnico di Torino} \\
Turin, Italy \\
\{giovanni.brignone, mihai.lazarescu, luciano.lavagno\}@polito.it}
}

\maketitle

\begin{abstract}
\Ac{HLS} enhances digital \acl{HW} design productivity through a 
high abstraction level. Even if the \ac{HLS} abstraction prevents 
fine-grained manual \ac{RTL} optimizations, it also enables 
automatable optimizations that would be unfeasible or hard to 
automate at \ac{RTL}. Specifically, we propose a task-level 
multi-pumping methodology to reduce resource utilization, 
particularly \acp{DSP}, while preserving the throughput of 
\ac{HLS} kernels modeled as \acp{DFG} targeting \aclp{FPGA}. The 
methodology exploits the \ac{HLS} resource sharing to 
automatically insert the logic for reusing the same functional 
unit for different operations. In addition, it relies on multi-clock \acp{DFG} 
to run the multi-pumped tasks at higher frequencies.
The methodology scales the pipeline \ac{II} and the clock frequency constraints
of resource-intensive tasks by a multi-pumping factor ($M$).
The looser \ac{II} allows sharing the 
same resource among $M$ different operations, while the tighter 
clock frequency preserves the throughput.
We verified that our methodology opens a new Pareto front in the 
throughput and resource space by applying it to open-source 
\ac{HLS} designs using \acl{SOTA} commercial \ac{HLS} and 
implementation tools by Xilinx.
The multi-pumped designs
require up to 40\% fewer \ac{DSP} resources at the same 
throughput as the original designs optimized for performance (i.e., running
at the maximum clock frequency) and
achieve up to 50\% better throughput using the same \acp{DSP}
as the original designs optimized for resources with a single clock.
\end{abstract}


\acresetall

\section{Introduction}
\label{sec:intro}

\Ac{HLS} raises the abstraction level of \acl{EDA} tools to improve the digital hardware designer's 
productivity. The high abstraction precludes some 
low-level manual optimizations, making the \ac{QoR} of \ac{HLS} 
circuits inferior to those manually optimized at the \ac{RTL}, 
especially for the area and maximum clock frequency 
\cite{fpga_hls_today}. On the other hand, we deem the \ac{HLS} 
description introduces new optimization opportunities at a high 
level.

We focus on \ac{HLS} designs modeled as 
\acp{DFG} (e.g., with \emph{dataflow} in Xilinx Vivado/Vitis 
\ac{HLS} \cite{vitis_hls}, \emph{hierarchy} in Siemens Catapult 
\ac{HLS} \cite{catapult}, or \emph{task functions} in Intel 
\ac{HLS} compiler \cite{intel_hls}).
Modeling \ac{HLS} designs as \acp{DFG} proved its effectiveness 
both in industrial \cite{vitis_libraries, finn} and academic 
\cites{mmm_hls, rosetta} projects.

\Iac{DFG} is a set of parallel computational tasks (C/C++ 
functions in \ac{HLS}) communicating asynchronously through 
\ac{FIFO} queues.
\Ac{HLS} tools typically implement \acp{DFG} as \acp{SCDFG}, 
where all the tasks share the same clock signal. Many modern 
\ac{HLS} tools do not support multi-clock designs 
\cite{vitis_hls, intel_hls}. Nevertheless, we can generalize 
\acp{SCDFG} to \acp{MCDFG} by assigning each task to a dedicated 
clock domain. The generalization enhances the tasks' flexibility 
and maximum frequency, limited only by the critical timing path 
local to the task rather than the global one. Clock architectures 
of modern \ac{FPGA} \acp{SoC} seamlessly support multiple clocks,
and the area overhead for safe \ac{CDC} is negligible since the 
tasks already communicate through \acp{FIFO}, which can be 
configured with independent read and write clocks with comparable 
resource utilization \cite{fifo_gen_pru}.

Multiple clock domains allow optimizations like multi-pumping, 
which reduces the area while preserving the throughput by reusing 
$M$ times a resource, usually \iac{DSP} unit in the \ac{FPGA} 
context, clocked at a frequency $M$ times larger than the rest of 
the system. Designers typically apply the technique at \ac{RTL} 
by manually inserting the custom logic to share the resource and 
safely perform \ac{CDC}.

In this work, we achieve multi-pumping at the task level by 
tuning only the high-level parameters of the tasks, in particular 
the pipeline \ac{II}, i.e., the clock cycles between the start of 
successive pipeline executions, and the clock constraint at the 
task granularity, taking advantage of the \ac{MCDFG}. The 
\ac{HLS} resource sharing algorithm automatically builds the 
logic for sharing the resource within a dataflow task. At the 
same time, the inter-task \acp{FIFO} allow safe \ac{CDC}. We 
focused on \acp{DSP}, which are critical in compute-intensive 
kernels and can run at high frequencies. However, the technique 
can multi-pump any shareable resource, including entire 
sub-functions.

\begin{figure}
\centering
\begin{tabular}{cc}%
\subfloat[\acs*{SCDFG}]{%
\hspace{1.03cm}
\begin{tikzpicture}[trim left=(read.west)]
\node (read) [rectangle, rounded corners, draw, color=myorange, fill=myorange!30, text=black, thick, inner xsep=-.2cm, minimum width=1.9cm, minimum height=.55cm, line width=1pt] {\footnotesize \texttt{ReadFromMem}};
	\node (fifo0) [draw, shape=rectangle split, rectangle split parts=3, fill=mygrey!20, minimum width=.25cm, minimum height=0, inner sep=-1.0pt, below=.15cm of read] {};
\node (window) [rectangle, rounded corners, draw, color=myorange, text=black, line width=1pt, fill=myorange!30, thick, below=.21cm of fifo0, inner xsep=-.2cm, minimum width=1.9cm, minimum height=.55cm] {\footnotesize \texttt{Window2D}};
	\node (fifo1) [draw, shape=rectangle split, rectangle split parts=3, fill=mygrey!20, minimum width=.25cm, minimum height=0, inner sep=-1pt, below=.15cm of window] {};
\node (filter) [rectangle, rounded corners, draw, color=myblue, text=black, line width=1pt, fill=myblue!30, thick, below=.21cm of fifo1, inner xsep=-.2cm, minimum width=1.9cm, minimum height=.55cm] {\footnotesize \texttt{Filter2D}};
	\node (fifo2) [draw, shape=rectangle split, rectangle split parts=3, fill=mygrey!20, minimum width=.25cm, minimum height=0, inner sep=-1pt, below=.15cm of filter] {};
\node (write) [rectangle, rounded corners, draw, color=myorange, text=black, line width=1pt, fill=myorange!30, thick, below=.21cm of fifo2, inner xsep=-.2cm, minimum width=1.9cm, minimum height=.55cm] {\footnotesize \texttt{WriteToMem}};
\coordinate[left=.2cm of $(window.west)!.5!(filter.west)$] (center_left);
\node (clk) [left=.2cm of center_left] {\footnotesize \shortstack{250\\\si{\mega\hertz}}};

\draw[thick] (read) -- (fifo0);
\draw[->, thick] (fifo0) -- (window);
\draw[thick] (window) -- (fifo1);
\draw[->, thick] (fifo1) -- (filter);
\draw[thick] (filter) -- (fifo2);
\draw[->, thick] (fifo2) -- (write);
\draw (clk) -- (center_left);
\draw[rounded corners] (center_left) |- (read.west);
\draw[rounded corners] (center_left) |- (write.west);
\draw (center_left |- window) |- (window.west);
\draw (center_left |- filter) |- (filter.west);
\end{tikzpicture}%
\label{subfig:conv2d_srdfg}%
}%
&%
\begin{tabular}[b]{c}%
\subfloat[\Acl{SOTA} \texttt{Filter2D}]{%
\hspace{.81cm}
\begin{tikzpicture}[trim left=(filter.west)]
\begin{pgfonlayer}{floating}
\node[draw, circle, fill=mypink!40, inner sep=0pt] (mul_0) {$\boldsymbol{\times}$};
\node[above=.2cm of mul_0.north west, xshift=-.05cm] (a_0) {};
\node[above=.2cm of mul_0.north east, xshift=.05cm] (b_0) {};
\node[below=.2cm of mul_0] (c_0) {};
\node[draw, circle, right=.15cm of mul_0, fill=mypink!40, inner sep=0pt] (mul_1) {$\boldsymbol{\times}$};
\node[above=.2cm of mul_1.north west, xshift=-.05cm] (a_1) {};
\node[above=.2cm of mul_1.north east, xshift=.05cm] (b_1) {};
\node[below=.2cm of mul_1] (c_1) {};
\node[right=.0cm of mul_1] (dots) {$\boldsymbol{\cdots}$};
\node[draw, circle, right=.0cm of dots, fill=mypink!40, inner sep=0pt] (mul_223) {$\boldsymbol{\times}$};
\node[above=.2cm of mul_223.north west, xshift=-.05cm] (a_223) {};
\node[above=.2cm of mul_223.north east, xshift=.05cm] (b_223) {};
\node[below=.2cm of mul_223] (c_223) {};
\node[draw, circle, right=.15cm of mul_223, fill=mypink!40, inner sep=0pt] (mul_224) {$\boldsymbol{\times}$};
\node[above=.2cm of mul_224.north west, xshift=-.05cm] (a_224) {};
\node[above=.2cm of mul_224.north east, xshift=.05cm] (b_224) {};
\node[below=.2cm of mul_224] (c_224) {};
\end{pgfonlayer}
\begin{pgfonlayer}{background}
	\node[draw, thick, rectangle, rounded corners, color=myblue, text=black, line width=1pt, fill=myblue!30, fit=(mul_0) (a_0) (c_0) (c_224) (b_224) (mul_224), minimum height=1.7cm, minimum width=3.16cm, inner sep=-10cm] (filter) {};
\end{pgfonlayer}
\begin{pgfonlayer}{floating}
\node[left=.2cm of filter] (clk) {\footnotesize \shortstack{250\\\si{\mega\hertz}}};
\end{pgfonlayer}

\draw (clk) -- (filter);
\draw[-{stealth[length=3pt, width=3pt]}] (a_0) -- (mul_0.north west);
\draw[-{stealth[length=3pt, width=3pt]}] (b_0) -- (mul_0.north east);
\draw[-{stealth[length=3pt, width=3pt]}] (mul_0) -- (c_0);
\draw[-{stealth[length=3pt, width=3pt]}] (a_1) -- (mul_1.north west);
\draw[-{stealth[length=3pt, width=3pt]}] (b_1) -- (mul_1.north east);
\draw[-{stealth[length=3pt, width=3pt]}] (mul_1) -- (c_1);
\draw[-{stealth[length=3pt, width=3pt]}] (a_223) -- (mul_223.north west);
\draw[-{stealth[length=3pt, width=3pt]}] (b_223) -- (mul_223.north east);
\draw[-{stealth[length=3pt, width=3pt]}] (mul_223) -- (c_223);
\draw[-{stealth[length=3pt, width=3pt]}] (a_224) -- (mul_224.north west);
\draw[-{stealth[length=3pt, width=3pt]}] (b_224) -- (mul_224.north east);
\draw[-{stealth[length=3pt, width=3pt]}] (mul_224) -- (c_224);
\end{tikzpicture}%
\label{subfig:filter2d_250}%
}%
\\%
\subfloat[Double-pumped \texttt{Filter2D}]{%
\hspace{.81cm}
\begin{tikzpicture}[trim left=(filter.west)]
\begin{pgfonlayer}{floating}
\node[draw, circle, fill=mypink!40, inner sep=0pt] (mul_0) {$\boldsymbol{\times}$};
\node[draw, trapezium, rotate=180, above=.3cm of mul_0.north west, xshift=.2cm, fill=mygrey!20, inner sep=2pt] (mux_a_0) {};
\node[above=.18cm of mux_a_0.south east, xshift=-.1cm] (a_0) {};
\node[above=.18cm of mux_a_0.south west, xshift=.1cm] (a_112) {};
\node[draw, trapezium, rotate=180, above=.3cm of mul_0.north east, xshift=-.2cm, fill=mygrey!20, inner sep=2pt] (mux_b_0) {};
\node[above=.18cm of mux_b_0.south east, xshift=-.1cm] (b_0) {};
\node[above=.18cm of mux_b_0.south west, xshift=.1cm] (b_112) {};
\node[draw, trapezium, below=.18cm of mul_0.south, fill=mygrey!20, inner sep=2pt] (demux_0) {};
\node[below=.18cm of demux_0.south west, xshift=-.1cm] (c_0) {};
\node[below=.18cm of demux_0.south east, xshift=.1cm] (c_112) {};
\node[right=.1cm of mul_0] (dots) {$\boldsymbol{\cdots}$};
\node[draw, circle, right=.1cm of dots, fill=mypink!40, inner sep=0pt] (mul_112) {$\boldsymbol{\times}$};
\node[draw, trapezium, rotate=180, above=.3cm of mul_112.north west, xshift=.2cm, fill=mygrey!20, inner sep=2pt] (mux_a_112) {};
\node[above=.18cm of mux_a_112.south east, xshift=-.1cm] (a_113) {};
\node[above=.18cm of mux_a_112.south west, xshift=.1cm] (a_224) {};
\node[draw, trapezium, rotate=180, above=.3cm of mul_112.north east, xshift=-.2cm, fill=mygrey!20, inner sep=2pt] (mux_b_112) {};
\node[above=.18cm of mux_b_112.south east, xshift=-.1cm] (b_113) {};
\node[above=.18cm of mux_b_112.south west, xshift=.1cm] (b_224) {};
\node[draw, trapezium, below=.18cm of mul_112.south, fill=mygrey!20, inner sep=2pt] (demux_113) {};
\node[below=.18cm of demux_113.south west, xshift=-.1cm] (c_113) {};
\node[below=.18cm of demux_113.south east, xshift=.1cm] (c_224) {};
\end{pgfonlayer}
\begin{pgfonlayer}{background}
\node[draw, thick, rectangle, rounded corners, color=myblue, text=black, line width=1pt, fill=myblue!30, fit=(a_0) (c_0) (c_112) (b_224), minimum height=1.7cm, minimum width=3.16cm, inner sep=-10cm] (filter) {};
\end{pgfonlayer}
\begin{pgfonlayer}{floating}
\node[left=.2cm of filter] (clk) {\footnotesize \shortstack{500\\\si{\mega\hertz}}};
\end{pgfonlayer}

\draw (clk) -- (filter);
\draw[-{stealth[length=3pt, width=3pt]}] (a_0) -- (mux_a_0.south east);
\draw[-{stealth[length=3pt, width=3pt]}] (a_112) -- (mux_a_0.south west);
\draw[-{stealth[length=3pt, width=3pt]}] (b_0) -- (mux_b_0.south east);
\draw[-{stealth[length=3pt, width=3pt]}] (b_112) -- (mux_b_0.south west);
\draw[-{stealth[length=3pt, width=3pt]}] (mux_a_0.north) -- (mul_0.north west);
\draw[-{stealth[length=3pt, width=3pt]}] (mux_b_0.north) -- (mul_0.north east);
\draw[-{stealth[length=3pt, width=3pt]}] (mux_a_112.north) -- (mul_112.north west);
\draw[-{stealth[length=3pt, width=3pt]}] (mux_b_112.north) -- (mul_112.north east);
\draw[-{stealth[length=3pt, width=3pt]}] (a_113) -- (mux_a_112.south east);
\draw[-{stealth[length=3pt, width=3pt]}] (a_224) -- (mux_a_112.south west);
\draw[-{stealth[length=3pt, width=3pt]}] (b_113) -- (mux_b_112.south east);
\draw[-{stealth[length=3pt, width=3pt]}] (b_224) -- (mux_b_112.south west);
\draw[-{stealth[length=3pt, width=3pt]}] (mul_0) -- (demux_0);
\draw[-{stealth[length=3pt, width=3pt]}] (demux_0.south west) -- (c_0);
\draw[-{stealth[length=3pt, width=3pt]}] (demux_0.south east) -- (c_112);
\draw[-{stealth[length=3pt, width=3pt]}] (mul_112) -- (demux_113);
\draw[-{stealth[length=3pt, width=3pt]}] (demux_113.south west) -- (c_113);
\draw[-{stealth[length=3pt, width=3pt]}] (demux_113.south east) -- (c_224);

\pgfresetboundingbox
\path (filter.north west) rectangle (filter.south east);

\end{tikzpicture}%
\label{subfig:filter2d_500}%
}%
\end{tabular}%
\end{tabular}
\caption{Task-level multi-pumping saves resources at equal 
throughput for \ac*{HLS} of \acfp*{DFG}. The \texttt{Filter2D} task 
from a 2D Convolution kernel 
\cite{xilinx_conv}~\protect\subref{subfig:conv2d_srdfg} is 
double-pumped~\protect\subref{subfig:filter2d_500} by doubling 
its clock frequency and \acs*{II} to save half of the multipliers 
of the single-clock 
solution~\protect\subref{subfig:filter2d_250}.}
\label{fig:conv2d}
\end{figure}
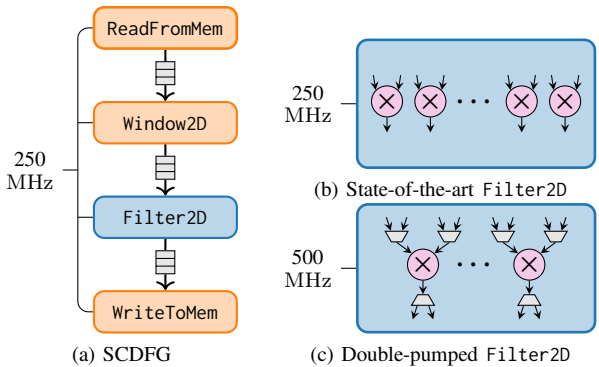

For example, consider a 2D Convolution \ac{HLS} kernel by 
\textcite{xilinx_conv}, implemented as \iac{SCDFG}, as shown in 
\cref{subfig:conv2d_srdfg},
where the rectangular nodes and the arrows represent the tasks 
and the \acp{FIFO}, respectively. At each iteration, the 
\texttt{Filter2D} task processes a convolution window of up to 
$15 \times 15$ elements, which requires computing \num{225}
\ac{MAC} operations bound to \acp{DSP}.
Thus, \iac{II} of \SI{1}{\cycle} requires \num{225} \acp{DSP}. On the other
hand, scaling the \ac{II} to 
\SI{2}{\cycles} implies that a new pipeline iteration starts every 
two clock cycles. Therefore, the pipeline has 
two cycles to compute the 225 operations. Hence, thanks to resource sharing, the \ac{HLS} binding
allocates only $\lceil 225/2\rceil = 113$ \acp{DSP}, each of which computes two \acp{MAC}.
Assume that we target a throughput of 
\SI{250}{\mega\sample\per\second}. With the \acl{SOTA} 
\ac{SCDFG} flow (\cref{subfig:filter2d_250}), we set the clock 
frequency of the whole \ac{DFG}, including the \texttt{Filter2D} 
task that allocates 225 \acp{DSP} at \SI{250}{\mega\hertz}. 
On the other hand, with our multi-pumping approach 
(\cref{subfig:filter2d_500}), we optimize the \texttt{Filter2D} 
task by scaling its \ac{II} to \SI{2}{\cycles}, to save half of 
the \acp{DSP}, and its clock frequency to 
\SI{500}{\mega\hertz}, to preserve the throughput.

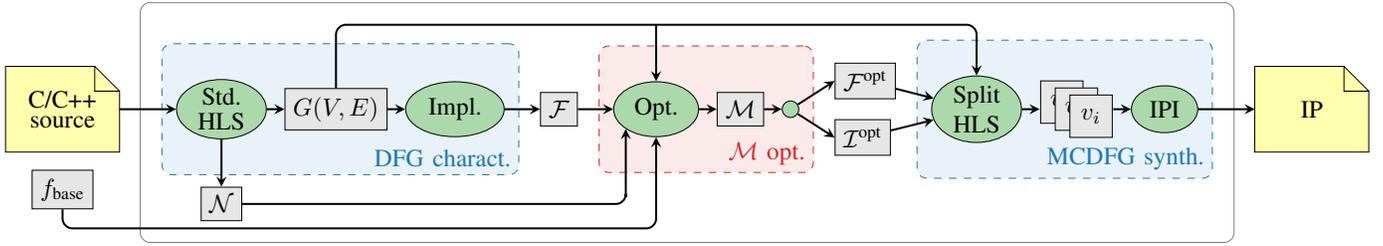
\begin{figure*}
\centering
\begin{tikzpicture}
\begin{pgfonlayer}{floating}
\node[draw, fill=mygreen!40, ellipse, inner ysep=0] (std_hls) {\shortstack{Std.\\HLS}};
\node[draw, fill=mygrey!20, rectangle, below=.67cm of std_hls] (ii) {$\mathcal{N}$};
\node[draw, fill=mygrey!20, rectangle, right=.25cm of std_hls] (dfg) {$G(V, E)$};
\node[draw, fill=mygreen!40, ellipse, right=.25cm of dfg] (impl_0) {Impl.};
\node[draw, fill=mygrey!20, rectangle, right=.5cm of impl_0] (f_max) {$\mathcal{F}$};
\node[draw, fill=mygreen!40, ellipse, right=.5cm of f_max] (opt) {Opt.};
\node[draw, fill=mygrey!20, rectangle, right=.25cm of opt] (mi) {$\mathcal{M}$};
\node[draw, fill=mygreen!40, circle, right=.25cm of mi, inner ysep=0] (comp_f_ii) {};
\node[draw, fill=mygrey!20, rectangle, right=.5cm of comp_f_ii, yshift=.4cm] (f_impl) {$\mathcal{F}^{\text{opt}}$};
\node[draw, fill=mygrey!20, rectangle, right=.5cm of comp_f_ii, yshift=-.4cm] (ii_impl) {$\mathcal{I}^{\text{opt}}$};
\node[draw, fill=mygreen!40, ellipse, right=.5cm of f_impl, inner ysep=0, yshift=-.4cm] (hls) {\shortstack{Split\\HLS}};
\node[right=.25cm of hls, inner sep=0] (v_hls_final) {\nodes};
\node[draw, fill=mygreen!40, ellipse, right=.25cm of v_hls_final] (ipi) {IPI};
\end{pgfonlayer}

\begin{pgfonlayer}{main}
\node[draw=myblue, dashed, fill=myblue!10, rectangle, rounded corners, fit=(std_hls) (impl_0), inner ysep=.5cm, inner xsep=.2cm] (param_extract) {};
\node[draw=myred, dashed, fill=myred!10, rectangle, rounded corners, fit=(opt) (comp_f_ii), inner ysep=.5cm, inner xsep=.2cm] (mi_opt) {};
\node[draw=myblue, dashed, fill=myblue!10, rectangle, rounded corners, fit=(hls) (v_hls_final) (ipi), inner ysep=.5cm, inner xsep=.2cm] (ip_synth) {};
\end{pgfonlayer}

\begin{pgfonlayer}{floating}
\node[anchor=south east] at (param_extract.south east) {\textcolor{myblue}{DFG charact.}};
\node[anchor=south east] at (mi_opt.south east) {\textcolor{myred}{$\mathcal{M}$ opt.}};
\node[anchor=south east] at (ip_synth.south east) {\textcolor{myblue}{MCDFG synth.}};

\draw[-stealth,thick] (std_hls) -- (ii);
\draw[-stealth,thick] (std_hls) -- (dfg);
\draw[-stealth,thick] (dfg) -- (impl_0);
\draw[-stealth,thick] (impl_0) -- (f_max);
\draw[-stealth,thick,rounded corners] (ii) -| (ii -| opt.south west) -- (opt.south west);
\draw[-stealth,thick] (f_max) -- (opt);
\coordinate[above=.8cm of opt] (above_opt);
\draw[-stealth,thick] (above_opt) -- (opt);
\draw[-stealth,thick] (opt) -- (mi);
\draw[-stealth,thick] (mi) -- (comp_f_ii);
\draw[-stealth,thick] (comp_f_ii) -- (f_impl.west);
\draw[-stealth,thick] (comp_f_ii) -- (ii_impl.west);
\draw[-stealth,thick] (f_impl) -- (hls);
\draw[-stealth,thick] (ii_impl) -- (hls);
\draw[-stealth,thick,rounded corners] (dfg) |- (above_opt) -| (hls);
\draw[-stealth,thick] (hls) -- (v_hls_final);
\draw[-stealth,thick] (v_hls_final) -- (ipi);
\end{pgfonlayer}

\begin{pgfonlayer}{background}
\node[draw=mygrey, rectangle, rounded corners, fit=(current bounding box), inner sep=.3cm] {};
\end{pgfonlayer}

\begin{pgfonlayer}{floating}
\node[draw, fill=myyellow!40, document, left=.8cm of std_hls, inner sep=0pt, minimum width=1.6cm, minimum height=1.2cm] (input_dfg) {\shortstack{C/C++\\source}};
\node[draw, fill=mygrey!20, rectangle, below=.25cm of input_dfg] (tput) {$f_{\text{base}}$};
\node[draw, fill=myyellow!40, document, right=.8cm of ipi, inner sep=0pt, minimum width=1.6cm, minimum height=1.2cm] (ip) {IP};

\draw[-stealth,thick,rounded corners] (tput) -- +(0, -.55) -| (opt);
\draw[-stealth,thick] (input_dfg) -- (std_hls);
\draw[-stealth,thick] (ipi) -- (ip);
\end{pgfonlayer}
\end{tikzpicture}
\caption[]{
Given the C/C++ source code of a \acf*{DFG} application and its base clock
frequency, the proposed workflow builds the optimized multi-pumped \acs*{IP} by
\begin{enumerate*}[label=(\alph*)]
\item analyzing the \ac*{DFG} (\emph{\ac*{DFG}~charact.}),
\item optimizing the multi-pumping factors (\emph{$\mathcal{M}$~opt.}),
and
\item synthesizing the multi-pumped \acs*{IP} (\emph{\ac*{MCDFG}~synth.}).
\end{enumerate*}
}
\label{fig:workflow}
\end{figure*}

This paper proposes an area-minimization methodology that preserves the
throughput via task-level multi-pumping for \ac{FPGA} \ac{HLS} designs
described as \acp{DFG}. Its effectiveness is validated on open-source designs
using the workflow shown in \cref{fig:workflow}, 
which generates an optimized multi-pumped \ac{IP} block from 
C/C++ source code using \acl{SOTA} \citeauthor{vivado} 
commercial tools \cite{vivado}.

To the best of our knowledge, this is the first work that combines multiple
clock domains with resource sharing in \ac{HLS} of \acp{DFG} for task-level
multi-pumping. The empirical results show that a new Pareto front opens in the
\ac{PPA} space, with circuits that use up to \SI{60}{\percent} fewer resources
at maximum throughput or achieve up to \SI{50}{\percent} higher throughput with
the same resources.

\section{Related work}
\label{sec:previous}

Our work is mainly related to \ac{QoR} improvement of \ac{HLS} designs by
tuning the \ac{HLS} directives (i.e., the instructions for the \ac{HLS}
compiler to control \acl{HW} optimizations such as loop pipelining), focusing
on multi-clock designs.

Several works \cite{autodse, multi_fidelity_hls, merlin_compiler,
model_based_hls, hiclockflow} optimize for performance the \ac{HLS} directives
applied to plain \acl{SW} code not intended for \ac{HLS} via \ac{DSE}.
However, the goals of their works differ from ours since we optimize for
resources while preserving the throughput of source code already optimized for
\ac{HLS}.
In addition, our methodology avoids time-consuming \acp{DSE} and analytically
computes the multi-pumping factor and, consequently, the corresponding \ac{II}
and clock frequency constraints.
Finally, they all consider only single clock designs, except for
\textcite{hiclockflow} (discussed further in \cref{subsec:prev_mrdfg}).

\Ac{HLS} design optimizations based on multiple clock domains work at the
\emph{operation level}, assigning domains at the low-level resource (e.g.,
adder or multiplier) granularity, typically during scheduling
\cite{low_power_multi_clock, multi_pump_hls, dsp_multi_pump}, or at the
\emph{task level}, assigning domains at function granularity (i.e.,
\acp{MCDFG}) \cite{legup_multi_clock, hiclockflow}.

\subsection{Operation-level multi-clock in high-level synthesis}
\label{subsec:prev_mrdfg}

\textcite{low_power_multi_clock} use multiple clock domains in \ac{HLS} to
reduce power consumption while preserving the throughput by halving the
operating frequency of two-cycle operations.
We instead focus on area and performance optimizations because power is only a
secondary quality metric for \ac{FPGA} designs after performance and area.

\textcite{multi_pump_hls} and \textcite{dsp_multi_pump} design 
double-pumped \ac{DSP} modules and use them in \ac{HLS} with 
custom resource-sharing algorithms. Theoretically,
\citeauthor{vivado} Vitis HLS supports double-pumped \ac{MAC} 
operations through user-callable functions from the 
\verb|dsp_builtins| library, but it is undocumented and faulty 
\cite{dsp_dp_vitis}. Our approach produces similar results when 
double-pumping a task. However, our task-level solution does not 
require custom modules, changes to the \ac{HLS} sharing 
algorithm, or changes to the source code. In addition, it can 
select multi-pumping factors greater than two, resulting in larger resource
savings.

\subsection{Task-level multi-clock in high-level synthesis}

\textcite{legup_multi_clock} focus on extending the LegUp 
\ac{HLS} tool to support \acp{MCDFG} synthesis but leave the 
selection of the clock frequencies to a suboptimal, 
time-consuming profiling-based approach. Our work focuses instead 
on a general methodology for exploiting the multiple clock domains. The 
workflow we define for building \acp{MCDFG}, based on 
\acl{SOTA} Xilinx tools, is just a means to apply our 
methodology.

\textcite{hiclockflow} propose \iac{DSE} methodology for 
maximizing the throughput under area constraints for \ac{HLS} of 
\ac{MCDFG} designs. They iteratively push for performance the 
\ac{HLS} loop directives applied to the bottleneck tasks. If a 
task is still a bottleneck after maximally pushing the directives 
(e.g., when the pipeline \ac{II} constraint is \SI{1}{\cycle}), they 
relax the directives of every task, increase the clock frequency 
of the bottleneck task, and restart the procedure.
The goal of our work is different since we minimize the area 
while preserving the throughput. The 
optimization approaches differ, too, since we optimize all the 
resource-intensive tasks independently of whether they are 
bottlenecks, and we never push the \ac{II} constraints, which 
the \ac{HLS} compiler may fail to meet (e.g., due to data dependencies).

\section{Background}

Given the \ac{DFG} throughput model defined in \cref{sec:dfg}, 
our multi-pumping methodology exploits the resource sharing 
executed by the \ac{HLS} binding step to build the sharing logic. 
The relaxed timing mode for the \ac{HLS} scheduling step ensures 
that the \ac{II} of the pipelines is independent of the target 
clock frequency, as explained in \cref{sec:hls}.

\subsection{Dataflow graph}
\label{sec:dfg}

\Iac{DFG} $G(V, E)$ is a set of tasks $v \in V$ running in 
parallel and communicating asynchronously through \ac{FIFO} 
channels $e \in E$.

In \ac{HLS}, each task is described as a C/C++ function whose 
core computational part typically consists of a pipelined loop. 
Given a task $v_i$ clocked at frequency $f_i$ and whose core loop 
is scheduled with \acl{II} $\mathit{II}_i$, an approximation of 
its throughput is
\begin{equation}
\Phi_i \coloneqq \frac{f_i}{\mathit{II}_i}.
\label{eqn:th_task}
\end{equation}
The maximum external \ac{DRAM} bandwidth can also limit 
throughput. However, this is out of the scope of our methodology 
since it does not change the overall throughput and, 
consequently, the \ac{DRAM} bandwidth requirements.

The overall \ac{DFG} throughput matches the one of the 
\emph{bottleneck task} (i.e., the task with the lowest 
throughput)
\begin{equation}
\Phi_G \coloneqq \min\limits_{v_i \in V} \Phi_i.
\label{eqn:th_net}
\end{equation}

According to \cref{eqn:th_task}, the high-level knobs for tuning 
the throughput of task $v_i$ are its clock frequency ($f_i$) and 
\acl{II} ($\mathit{II}_i$). All tasks share the same clock in 
\acfp{SCDFG}. Thus, $f_i$ is the same for all tasks and is 
bounded by the global (i.e., among all tasks) critical path. 
Therefore, only the $\mathit{II}_i$ can be tuned independently 
for each task. In \iac{MCDFG}, on the other hand, the clock 
frequency can be set individually for each task. This additional 
degree of freedom allows for higher flexibility and tasks 
frequencies than \ac{SCDFG} since the clock frequency of a task 
is limited only by its local critical path and not the one of the 
whole \ac{DFG}.

\subsection{High-level synthesis}
\label{sec:hls}

Our multi-pumping technique relies on two key concepts of the 
\ac{HLS} tools:
\begin{enumerate*}[label=(\arabic*)]
\item the minimum pipeline \ac{II} is independent of the clock 
frequency constraint when the scheduler works in \emph{relaxed 
timing} mode (i.e., the clock frequency is subordinate to meet 
the \ac{II} constraints), and
\item the level of \emph{resource sharing} is directly dependent 
on the \ac{II}.
\end{enumerate*}
Both are implemented by the \ac{HLS} back-end that generates the 
\acl{HW} description. The timing model is used during 
\emph{scheduling} and the resource sharing is done during 
\emph{binding} \cite{vitis_hls}.

\subsubsection{Scheduling}

Scheduling assigns operations to specific clock cycles; thus, it 
also implements loop and function \emph{pipelining}. Designers 
can constrain the \ac{II} of the pipelines, which is lower bound
by the resource constraints and the data dependencies.
Consider the \ac{DDG} modeling the data dependencies in a 
kernel. Given a cycle $\theta$ in the \ac{DDG}, we define 
$\mathit{delay}_\theta$ as the sum of the delays of the operations 
along $\theta$ and 
$\mathit{dist}_\theta$ as the total loop-carried dependence 
distance along $\theta$. The lower bound of the \ac{II} is 
\begin{equation}
\mathit{II}^{\text{min}}
\coloneqq \max\limits_{\theta \in \mathrm{DDG}}\left(\frac{\mathit{delay}_\theta}{\mathit{dist}_\theta}\right).
\label{eqn:ii_min}
\end{equation}
The associated cycle is called \emph{critical} 
\cite{sw_pipeline}.

For example, consider the following loop to be scheduled:
\begin{center}
\begin{minipage}{.66\linewidth}
\begin{tcolorbox}[standard jigsaw, opacityback=0]
\begin{minted}{c++}
for (int i = 0; i < N; i++)
    a = a + b;
\end{minted}
\end{tcolorbox}
\end{minipage}
\end{center}
The read-after-write dependency on \texttt{a},
produced at the $i$-th iteration and consumed at the $i+1$-th iteration,
introduces a cycle $\theta$ in the \ac{DDG}.
$\mathit{delay}_\theta$ is the latency of the adder computing \texttt{a+b}.
$\textit{dist}_\theta$ is 1 since \texttt{a} is consumed at the iteration after it is produced.
Therefore, \cref{eqn:ii_min} implies that the minimum \ac{II} for this loop equals the latency of the adder.

The clock constraints determine how many operations fit within 
a clock cycle, thus affecting the depth of the pipelines.
The pipeline depth determines the latencies of its operations,
impacting the critical cycle and, in turn, the \ac{II} lower bound. However, 
the \ac{II} constraints take precedence over clock constraints in 
\emph{relaxed timing} mode, yielding lower \ac{II} pipelines in 
exchange for potential \ac{HLS} timing violations. These are 
usually acceptable at \ac{HLS} time since \ac{HLS} timing estimations may be 
overly pessimistic \cite{fpga_hls_today}, and downstream 
implementation steps may resolve them.

\subsubsection{Binding}
\label{subsec:bind}

Binding assigns each operation to a compatible functional unit, 
depending on resource and performance (e.g., clock frequency, \ac{II}) 
constraints.

\emph{Resource sharing} is a crucial binding optimization that 
maps operations of the same type to the same functional unit, 
scheduled on different clock cycles or under mutually exclusive 
conditions (e.g., on different if-then-else branches). The 
\ac{II} constraints directly affect the degree of resource 
sharing. In particular, if a pipeline scheduled with \iac{II} of 
\SI[parse-numbers=false]{\text{$\mathit{\acs{II}}_i$}}{\cycles} 
computes $N^{\text{\acs{OP}}}_i$ \acfp{OP} of the same kind at 
each iteration, the binding step allocates 
$N^{\text{\acs{FU}}}_i$ \acfp{FU}, with
\begin{equation}
N^{\text{\acs{FU}}}_i
\coloneqq \left\lceil \frac{N^{\text{\acs{OP}}}_i}{\mathit{\acs{II}}_i} \right\rceil.
\label{eqn:n_fu}
\end{equation}

Note that the \aclp{OP} can be either computations or memory 
accesses. The \aclp{FU} associated with the memory \aclp{OP} are 
ports proportional to the partitioning factors (i.e., 
the number of submemories into which a memory resource is divided 
to increase its parallelism). Therefore, larger \ac{II} values 
result in fewer \aclp{FU} and smaller memory partitioning 
factors.

\begin{figure}
\centering
\subfloat[][Source code]{\includestandalone[width=\linewidth]{figures/conv2d-src}\label{subfig:filter_src}}

\subfloat[][Execution with \ac{II} \num{1} cycle]{
\hspace{1cm}
\begin{tikzpicture}[trim left=(origin), every node/.style={minimum width=.7cm, inner sep=1pt}]
\foreach \x in {0, ..., 3}{
\node[draw, thick, circle, fill=mygreen!30] (read_\x) at (1.5 + 4 * \x, 0 - \x) {\texttt{rd}};

\node[draw, thick, circle, fill=mypink!40] (m_0_\x) at (0 + 4 * \x, -1 - \x) {$\boldsymbol{\times}$};
\node[draw, thick, circle, fill=mypink!40] (m_1_\x) at (1 + 4 * \x, -1 - \x) {$\boldsymbol{\times}$};

\draw[->, thick] (read_\x) -- (m_0_\x);
\draw[->, thick] (read_\x) -- (m_1_\x);

\node[draw, thick, circle, fill=myorange!30] (s_0_\x) at (.5 + 4 * \x, -2 - \x) {$\boldsymbol{+}$};
\node[draw, thick, circle, fill=mypink!40] (m_2_\x) at (2 + 4 * \x, -2 - \x) {$\boldsymbol{\times}$};
\node[draw, thick, circle, fill=mypink!40] (m_3_\x) at (3 + 4 * \x, -2 - \x) {$\boldsymbol{\times}$};

\draw[->, thick] (m_0_\x) -- (s_0_\x);
\draw[->, thick] (m_1_\x) -- (s_0_\x);
\draw[->, thick] (read_\x) -- (m_2_\x);
\draw[->, thick] (read_\x) -- (m_3_\x);

\node[draw, thick, circle, fill=myorange!30] (s_1_\x) at (2.5 + 4 * \x, -3 - \x) {$\boldsymbol{+}$};

\draw[->, thick] (m_2_\x) -- (s_1_\x);
\draw[->, thick] (m_3_\x) -- (s_1_\x);

\node[draw, thick, circle, fill=myorange!30] (s_2_\x) at (1.5 + 4 * \x, -4 - \x) {$\boldsymbol{+}$};

\draw[->, thick] (s_0_\x) -- (s_2_\x);
\draw[->, thick] (s_1_\x) -- (s_2_\x);

\node[draw, thick, circle, fill=myblue!30] (write_\x) at (1.5 + 4 * \x, -5 - \x) {\texttt{wr}};

\draw[->, thick] (s_2_\x) -- (write_\x);
}

%

\foreach \x in {-6, ..., 0}{
\coordinate (step\x) at (0, \x + .5);
\ifthenelse{\x < 0}{
\draw[dashed] (current bounding box.west |- step\x) -- (current bounding box.east |- step\x);
}{
\path (current bounding box.west |- step\x) -- (current bounding box.east |- step\x);
}
}

\foreach \x in {-1, ..., 1}{
\coordinate (vert\x) at (3.5 + 4 * \x, 0);
\ifthenelse{\x>-1}{
\draw[dotted] (vert\x |- current bounding box.north) -- (vert\x |- current bounding box.south);
}{
\path (vert\x |- current bounding box.north) -- (vert\x |- current bounding box.south);
}
}

\foreach \x in {1, ..., 3}{
	\pgfmathtruncatemacro\y{\x-1}
	\node[rectangle, rounded corners, draw=red, ultra thick, fit=(m_2_\y) (m_3_\y) (m_0_\x) (m_1_\x)] (m_fit_\x) {};
}


\draw[red, |-|, ultra thick] (5.5, 0.4) -- (5.5, -.4) node [midway, right, xshift=5pt] {\textbf{II = 1 cycle}};

\pgfresetboundingbox
\path (step0.north -| vert-1.west) rectangle ($(write_1.south -| m_3_2.east) + (1.5pt, 0)$);

\coordinate (origin) at (current bounding box.north west);
\draw[->, >={Stealth[length=3mm, width=3mm]}, thick] (origin) -- node[midway, rotate=90, anchor=south, yshift=.3cm] {\Large Time (cycles)} (current bounding box.south west);
\draw[->, >={Stealth[length=3mm, width=3mm]}, thick] (origin) -- node[midway, above=.2cm] {\Large Loop iteration (\#)} (current bounding box.north east);


\end{tikzpicture}
\label{subfig:filter_ii_1}
}

\subfloat[][Execution with \ac{II} \num{2} cycles]{
\hspace{1cm}
\begin{tikzpicture}[trim left=(origin), every node/.style={minimum width=.7cm, inner sep=1pt}]
\foreach \x in {0, ..., 2}{

\node[draw, thick, circle, fill=mygreen!30] (read_\x) at (1.5 + 4 * \x, 0 - 2 * \x) {\texttt{rd}};

\node[draw, thick, circle, fill=mypink!40] (m_0_\x) at (0 + 4 * \x, -1 - 2 * \x) {$\boldsymbol{\times}$};
\node[draw, thick, circle, fill=mypink!40] (m_1_\x) at (1 + 4 * \x, -1 - 2 * \x) {$\boldsymbol{\times}$};

\node[draw, thick, circle, fill=myorange!30] (s_0_\x) at (.5 + 4 * \x, -2 - 2 * \x) {$\boldsymbol{+}$};
\node[draw, thick, circle, fill=mypink!40] (m_2_\x) at (2 + 4 * \x, -2 - 2 * \x) {$\boldsymbol{\times}$};
\node[draw, thick, circle, fill=mypink!40] (m_3_\x) at (3 + 4 * \x, -2 - 2 * \x) {$\boldsymbol{\times}$};

\node[draw, thick, circle, fill=myorange!30] (s_1_\x) at (2.5 + 4 * \x, -3 - 2 * \x) {$\boldsymbol{+}$};

\node[draw, thick, circle, fill=myorange!30] (s_2_\x) at (1.5 + 4 * \x, -4 - 2 * \x) {$\boldsymbol{+}$};
\node[draw, thick, circle, fill=myblue!30] (write_\x) at (1.5 + 4 * \x, -5 - 2 * \x) {\texttt{wr}};

\draw[->, thick] (read_\x) -- (m_0_\x);
\draw[->, thick] (read_\x) -- (m_1_\x);
\draw[->, thick] (m_0_\x) -- (s_0_\x);
\draw[->, thick] (m_1_\x) -- (s_0_\x);
\draw[->, thick] (read_\x) -- (m_2_\x);
\draw[->, thick] (read_\x) -- (m_3_\x);
\draw[->, thick] (m_2_\x) -- (s_1_\x);
\draw[->, thick] (m_3_\x) -- (s_1_\x);
\draw[->, thick] (s_0_\x) -- (s_2_\x);
\draw[->, thick] (s_1_\x) -- (s_2_\x);
\draw[->, thick] (s_2_\x) -- (write_\x);
}

%

\foreach \x in {-6, ..., 0}{
\coordinate (step\x) at (0, \x + .5);
\ifthenelse{\x < 0}{
\draw[dashed] (current bounding box.west |- step\x) -- (current bounding box.east |- step\x);
}{
\path (current bounding box.west |- step\x) -- (current bounding box.east |- step\x);
}
}

\foreach \x in {-1, ..., 1}{
\coordinate (vert\x) at (3.5 + 4 * \x, 0);
\ifthenelse{\x>-1}{
\draw[dotted] (vert\x |- current bounding box.north) -- (vert\x |- current bounding box.south);
}{
\path (vert\x |- current bounding box.north) -- (vert\x |- current bounding box.south);
}
}

\foreach \x in {0, ..., 2}{
	\foreach \y in {0, 1}{
		\pgfmathtruncatemacro\z{\y*2}
		\pgfmathtruncatemacro\w{\z+1}
		\node[rectangle, rounded corners, draw=red, ultra thick, fit=(m_\z_\x) (m_\w_\x)] (m_fit_\x_\y) {};
	}
}


\draw[red, |-|, ultra thick] (5.5, 0.4) -- (5.5, -1.4) node [midway, right, xshift=5pt, yshift=5pt] {\textbf{II = 2 cycles}};

\pgfresetboundingbox
\path (step0.north -| vert-1.west) rectangle ($(m_fit_2_1.south -| m_3_2.east) + (1.5pt, 0)$);

\coordinate (origin) at (current bounding box.north west);
\draw[->, >={Stealth[length=3mm, width=3mm]}, thick] (origin) -- node[midway, rotate=90, anchor=south, yshift=.3cm] {\Large Time (cycles)} (current bounding box.south west);
\draw[->, >={Stealth[length=3mm, width=3mm]}, thick] (origin) -- node[midway, above=.2cm] {\Large Loop iteration (\#)} (current bounding box.north east);

\end{tikzpicture}
\label{subfig:filter_ii_2}
}
\caption{
The pipeline \acf*{II} directly affects the resource sharing. 
For example, in the \texttt{Filter2D} task 
\protect\subref{subfig:filter_src}, the pipeline with 
$\mathit{\acs*{II}} = \SI{1}{\cycle}$ \protect\subref{subfig:filter_ii_1}
computes four multiplications per clock cycle in steady state, 
while the one with $\mathit{\acs*{II}} = \SI{2}{\cycles}$
\protect\subref{subfig:filter_ii_2} only two. Thus, the latter 
datapath allocates half of the multipliers.
}
\label{fig:res_vs_freq-ii}
\end{figure}
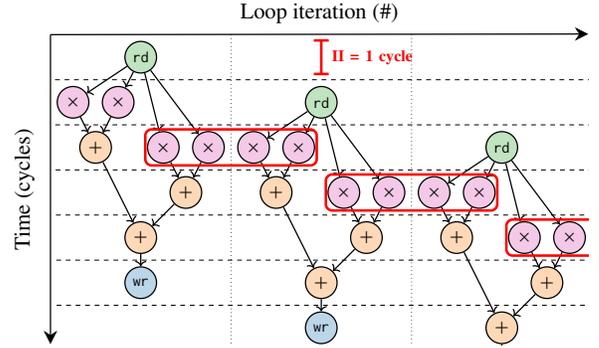
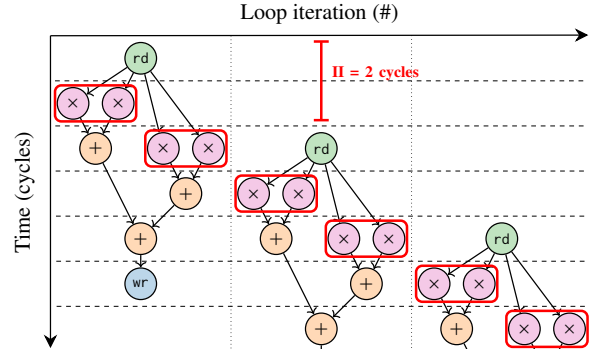

Consider the \texttt{Filter2D} task from the 2D Convolution 
kernel introduced in \cref{sec:intro}, whose source code is in 
\cref{subfig:filter_src}.
Assuming a filter of size $2 \times 2$ (i.e., 
$\text{\texttt{FILTER\_V\_SIZE}} = 
\text{\texttt{FILTER\_H\_SIZE}} = 2$), 
with the schedule with \iac{II} of \SI{1}{\cycle} (shown in 
\cref{subfig:filter_ii_1}, where the nodes represent the operations, and the edges 
the data dependencies), at the steady-state, four 
multiplications are computed in parallel on different data within 
the same clock cycle (highlighted by the red rectangle), thus requiring four \ac{DSP}-mapped multipliers.
With \iac{II} of \SI{2}{\cycles} instead (see 
\cref{subfig:filter_ii_2}), only two multiplications are computed 
per clock cycle. Therefore, the binding step allocates only two multipliers
and shares these among two multiplications.

\section{Task-level multi-pumping}
\label{sec:algo}

We multi-pump the resources of task $v_i$ by simultaneously 
scaling by a multi-pumping factor $M_i$ the \ac{II} and the clock 
frequency of $v_i$.

The underlying principles of our approach are:
\begin{itemize}
\item \Cref{eqn:th_net} allows tuning each task independently 
without reducing the overall \ac{DFG} throughput, as long as the 
throughput of the task does not get lower than the bottleneck 
task one.
\item As discussed in \cref{subsec:bind}, scaling the \ac{II} of 
a pipelined loop by a factor $M_i$ allows reusing the same 
functional unit for $M_i$ operations in different clock cycles.
\item \Cref{eqn:th_task} implies that the task throughput is 
unchanged if we scale by $M_i$ the task clock frequency together 
with the \ac{II}.
\end{itemize}

Assume that $v_i$ meets the timing constraints up to $f^{\text{max}}_i$ and 
computes $N^{\text{\acs{OP}}}_i$ \aclp{OP} mapped to \acp{DSP}.
Moreover, the non-multi-pumped tasks are 
clocked at $f_{\text{base}}$ (i.e., the clock constraint given by 
the designer). The maximum multi-pumping factor for task $v_i$ is
\begin{equation}
M^{\text{max}}_i
\coloneqq \min\left(\left\lfloor\frac{f^{\text{max}}_i}{f_{\text{base}}}\right\rfloor, N^{\text{\acs{OP}}}_i\right).
\label{eqn:m_opt}
\end{equation}

It is worth noting that our task-level multi-pumping 
\emph{changes only the \ac{HLS} directives while using the 
\ac{HLS} tool as a black box and without requiring manual source 
code restructuring}. The automation of this step will be the subject of future work.

\section{Multi-pumping workflow}
\label{sec:workflow}

To validate our task-level multi-pumping, we define a workflow from 
the C/C++ source code to an optimized \ac{MCDFG} \ac{IP} block
compatible with \citeauthor{vivado} tools \cite{vivado}, as shown 
in \cref{fig:workflow}. The main steps of the workflow are
\begin{enumerate*}[label=(\Alph*)]
\item \emph{\ac{DFG} characterization} to extract the maximum clock frequency
and the number of \ac{DSP} operations of each task, needed by the later steps,
\item \emph{multi-pumping factor optimization} to select the 
multi-pumping factor of each task, and
\item \emph{\ac{MCDFG} synthesis} to generate the multi-pumped 
\ac{IP}.
\end{enumerate*}

\subsection{Dataflow graph characterization}
\label{subsec:dfg_char}

For each task in the \ac{DFG} $G(V,E)$, we collect the number of 
\ac{DSP} \aclp{OP} ($\mathcal{N} = \{N^{\text{\acs{OP}}}_i, 
\forall v_i \in V\}$) from the reports of the standard \ac{SCDFG} \ac{HLS}.
We collect the maximum frequency meeting the timing constraints ($\mathcal{F} =
\{f^{\text{max}}_i, \forall v_i \in V\}$) from the post-implementation reports
of the \ac{SCDFG}.
We execute the implementation with a tight clock constraint (e.g.,
\SI{500}{\mega\hertz}) and at the lowest pipeline \ac{II}, which is the
worst case for the critical cycle (defined in \cref{sec:hls}).
Indeed, when multi-pumping increases the \ac{II}, it relaxes the critical
cycle, allowing deeper pipelines and shorter critical paths, thus higher clock
frequencies.

We do not extract $\mathcal{F}$ from the earlier-available \ac{HLS} clock
frequency estimations since they are unreliable \cite{fpga_hls_today}.
We run the \ac{SCDFG} implementation only once, so the overhead is usually
acceptable.
However, if a fast flow is required (e.g., in early design phases), we can run
only the logic synthesis step without placement and routing. The timing
estimations at the logic synthesis step are more accurate than the one of the
\ac{HLS} compiler since they have access to lower-level information.
When the estimated maximum frequency is less than the actual one, we miss
chances of saving resources because of lower multi-pumping factors, as per
\cref{eqn:m_opt}.
On the contrary, if the frequency is overestimated, the timing fails during
implementation.

\subsection{Multi-pumping factor optimization}

We select the multi-pumping factors ($\mathcal{M} = \{M_i, 
\forall v_i \in V\}$) that minimize the \ac{DSP} utilization. If 
$v_i$ contains \aclp{OP} mapped to \ac{DSP}, we set $M_i = 
M^{\text{max}}_i$, as defined by \cref{eqn:m_opt}. Otherwise, we 
do not apply multi-pumping to $v_i$. 

\subsection{Multi-clock dataflow graph synthesis}

Xilinx Vitis HLS cannot synthesize \acp{MCDFG} directly since
it supports only one clock domain per the design. However, the \texttt{dataflow} 
directive generates several independent 
modules, one for each task, and interconnects them in a top-level 
module. Thus, we run a \emph{split} \ac{HLS}, synthesizing 
each task separately (i.e., setting it as the top module) with 
its clock constraint.

The Xilinx \ac{HLS} binding algorithm guarantees optimal 
resource sharing if guided by resource constraints only. 
Therefore, we constrain the number of \acp{DSP} according to 
\cref{eqn:n_fu}. For instance, if we multi-pump with a factor 
$M_i$ a task $v_i$ that originally uses $N^{\text{\acs{DSP}}}_i$, 
we constrain its \acp{DSP} to $\left\lceil 
N^{\text{\acs{DSP}}}_i/M_i \right\rceil$.

In principle, we could also scale down the memory partitioning factors by $M_i$
to reduce on-chip memory resource usage, namely \acp{BRAM} and registers.
However, we cannot apply this optimization to the test cases considered in
\cref{sec:bench} with Xilinx \ac{HLS}. Indeed, the tool ignores the coarser
partitioning directives and automatically partitions the memories, presumably
to minimize the pipelines \ac{II}, regardless of the provided directives. We
plan to revisit the issue as a more recent version of the \ac{HLS} tool is
available.

Finally, we interconnect the tasks synthesized separately using 
the Vivado \ac{IPI}.

The Xilinx \ac{HLS} tools use \acp{FIFO} as inter-task communication channels 
when data are produced and consumed in the same order; otherwise, 
\acl{PIPO} buffers. Our method could support both, but 
since the Xilinx \ac{IPI} flow does not provide a configurable 
multi-clock \acl{PIPO} buffer, we currently only support \ac{FIFO} 
channels using the \citeauthor{vivado} \ac{FIFO} generator 
\cite{fifo_gen}. \acp{FIFO} are configured with independent 
clocks for read and write ports when interconnecting tasks 
assigned to different clock domains.

\section{Evaluation}
\label{sec:bench}

We verify the applicability and the advantages in the \ac{PPA} space of our
task-level multi-pumping workflow, described in \cref{sec:workflow}, by
applying it to open-source \ac{HLS} designs.

Our experiments target the embedded platform Zynq UltraScale+ 
\ac{FPGA} \ac{SoC} hosted by the Avnet Ultra96v1 board 
\cite{ultra96v1}. We use Vitis HLS 2022.2 \cite{vitis_hls} and 
Vivado HLS 2019.2 \cite{vivado_hls} for the synthesis and Vivado 
2022.2 \cite{vivado} for the implementation.

We collect the resource utilization from the post-implementation reports and
the power estimations from the post-implementation static power analysis.
We verify that the throughput (i.e., the number of output samples produced in
the unit of time) matches the theoretical one by measuring the time for
\num{10000} executions in auto-restart mode \cite{vitis_hls} (to make the time
overhead for control negligible) of the kernels in \acl{HW}, using the PYNQ
\aclp{API} \cite{pynq}.

We apply our flow to some open-source \ac{HLS} designs, including
\begin{enumerate*}[label=(\alph*)]
\item the \emph{2D Convolution} from the Vitis 
Tutorials \cite{xilinx_conv} already introduced in \cref{sec:intro},
\item the \emph{Optical Flow} from the Rosetta 
suite \cite{rosetta}, and
\item the \emph{\ac{VMS}} \cite{vms}, a drug discovery accelerator.
\end{enumerate*}

For each design, we compare the multi-pumped implementations (\emph{M-Pump})
with the original ones (\emph{Base}) and with the best \ac{SCDFG}
implementations without source code changes (\emph{S-Pump}).
For the \emph{S-Pump} implementations, we apply our flow without the
generalization to \ac{MCDFG}. Thus, if task $v_i$ is ``single-pumped'' by a
factor $S_i$, we scale by $S_i$ its \ac{II}, as with our original workflow, and
the clock frequency of the whole kernel.
The maximum ``single-pumping'' factor for each task is lower than the
corresponding maximum multi-pumping factor (defined by \cref{eqn:m_opt}) since
it is at most
\begin{equation}
S^{\text{max}}_i
\coloneqq \left\lfloor \dfrac{\min\limits_{\forall v_i \in V} f_i^{\text{max}}}{f_{\text{base}}} \right\rfloor.
\end{equation}

\begin{figure*}
\centering
\phantom{}\hfill%
\subfloat[2D Convolution]{%
\centering%
\newcommand{\fileName}{conv}%
\newcommand{\legendPos}{north east}%
\newcommand{\memBound}{inf}%
\newcommand{\degSBound}{167}%
\newcommand{\degMBound}{251}%
\newcommand{\implBase}{(165,62.5) (250,62.5)}%
\newcommand{\implSPump}{(165,31.38888888888889)}%
\newcommand{\implMPump}{(165,20.833333333333336) (250,31.38888888888889)}%
\hspace{.73cm}
\begin{tikzpicture}[trim axis left, trim axis right]
\begin{axis}[
xlabel={Throughput (\si{\mega\sample\per\second})},
xlabel style={
yshift=2pt,
font=\footnotesize,
},
ylabel={DSP utilization (\si{\percent})},
ylabel style={
yshift=-8pt,
font=\footnotesize,
},
tick label style={
font=\footnotesize,
},
grid=major,
grid style={densely dotted},
legend entries={\emph{Base}, \emph{S-Pump}, \emph{M-Pump}},
width=.326\textwidth,
height=.326\textwidth,
legend style={
font=\footnotesize,
legend pos=\legendPos,
inner xsep=2pt,
inner ysep=.5pt,
rounded corners,
},
ymin=0,
ymax=100,
xmin=0,
set layers,
]
\pgfplotstableread[col sep=comma]{data/\fileName.csv}\datatable

\addplot[const plot, myblue, on layer=axis foreground, very thick, restrict x to domain=0:\memBound, forget plot] table [x=throughput, y=dsp_base] {\datatable};
\addplot[const plot, myblue, on layer=axis foreground, very thick, dashed, forget plot, restrict x to domain=\memBound:inf] table [x=throughput, y=dsp_base] {\datatable};
\addplot[only marks, mark options={fill=myblue, draw=myblue, scale=1.5}, forget plot] coordinates {\implBase};
\addplot[const plot, myorange, very thick, restrict x to domain=0:\degSBound, restrict x to domain=0:\memBound, forget plot] table [x=throughput, y=dsp_s-pump] {\datatable};
\addplot[const plot, myorange, very thick, dashed, forget plot, restrict x to domain=0:\degSBound, restrict x to domain=\memBound:inf] table [x=throughput, y=dsp_s-pump] {\datatable};
\addplot[only marks, mark options={fill=myorange, draw=myorange, scale=1.5}, forget plot] coordinates {\implSPump};
\addplot[const plot, mygreen, very thick, restrict x to domain=0:\degMBound, restrict x to domain=0:\memBound, forget plot] table [x=throughput, y=dsp_m-pump] {\datatable};
\addplot[const plot, mygreen, very thick, dashed, forget plot, restrict x to domain=0:\degMBound, restrict x to domain=\memBound:inf] table [x=throughput, y=dsp_m-pump] {\datatable};
\addplot[only marks, mark options={fill=mygreen, draw=mygreen, scale=1.5}, forget plot] coordinates {\implMPump};

\addplot[myblue, very thick] coordinates {(-1, -1)};
\addplot[myorange, very thick] coordinates {(-1, -1)};
\addplot[mygreen, very thick] coordinates {(-1, -1)};
\end{axis}
\end{tikzpicture}%
\label{subfig:conv2d_tp_vs_mult}%
}%
\hfill%
\hfill%
\hfill%
\subfloat[Optical Flow]{%
\centering%
\newcommand{\fileName}{optical}%
\newcommand{\legendPos}{north east}%
\newcommand{\memBound}{175}%
\newcommand{\degSBound}{151}%
\newcommand{\degMBound}{227}%
\newcommand{\implBase}{(150,52.22222222222222) (175,52.22222222222222)}%
\newcommand{\implSPump}{(150,26.666666666666668)}%
\newcommand{\implMPump}{(150,18.333333333333336) (175,29.444444444444443)}%
\hspace{.73cm}
\begin{tikzpicture}[trim axis left, trim axis right]
\begin{axis}[
xlabel={Throughput (\si{\mega\sample\per\second})},
xlabel style={
yshift=2pt,
font=\footnotesize,
},
ylabel={DSP utilization (\si{\percent})},
ylabel style={
yshift=-8pt,
font=\footnotesize,
},
tick label style={
font=\footnotesize,
},
grid=major,
grid style={densely dotted},
legend entries={\emph{Base}, \emph{S-Pump}, \emph{M-Pump}},
width=.326\textwidth,
height=.326\textwidth,
legend style={
font=\footnotesize,
legend pos=\legendPos,
inner xsep=2pt,
inner ysep=.5pt,
rounded corners,
},
ymin=0,
ymax=100,
xmin=0,
set layers,
]
\pgfplotstableread[col sep=comma]{data/\fileName.csv}\datatable

\addplot[const plot, myblue, on layer=axis foreground, very thick, restrict x to domain=0:\memBound, forget plot] table [x=throughput, y=dsp_base] {\datatable};
\addplot[const plot, myblue, on layer=axis foreground, very thick, dashed, forget plot, restrict x to domain=\memBound:inf] table [x=throughput, y=dsp_base] {\datatable};
\addplot[only marks, mark options={fill=myblue, draw=myblue, scale=1.5}, forget plot] coordinates {\implBase};
\addplot[const plot, myorange, very thick, restrict x to domain=0:\degSBound, restrict x to domain=0:\memBound, forget plot] table [x=throughput, y=dsp_s-pump] {\datatable};
\addplot[const plot, myorange, very thick, dashed, forget plot, restrict x to domain=0:\degSBound, restrict x to domain=\memBound:inf] table [x=throughput, y=dsp_s-pump] {\datatable};
\addplot[only marks, mark options={fill=myorange, draw=myorange, scale=1.5}, forget plot] coordinates {\implSPump};
\addplot[const plot, mygreen, very thick, restrict x to domain=0:\degMBound, restrict x to domain=0:\memBound, forget plot] table [x=throughput, y=dsp_m-pump] {\datatable};
\addplot[const plot, mygreen, very thick, dashed, forget plot, restrict x to domain=0:\degMBound, restrict x to domain=\memBound:inf] table [x=throughput, y=dsp_m-pump] {\datatable};
\addplot[only marks, mark options={fill=mygreen, draw=mygreen, scale=1.5}, forget plot] coordinates {\implMPump};

\addplot[myblue, very thick] coordinates {(-1, -1)};
\addplot[myorange, very thick] coordinates {(-1, -1)};
\addplot[mygreen, very thick] coordinates {(-1, -1)};
\end{axis}
\end{tikzpicture}%
\label{subfig:optical_tp_vs_dsp}%
}%
\hfill%
\hfill%
\hfill%
\subfloat[VMS]{%
\centering%
\newcommand{\fileName}{vms}%
\newcommand{\legendPos}{south east}%
\newcommand{\memBound}{inf}%
\newcommand{\degSBound}{27.75}%
\newcommand{\degMBound}{41.5}%
\newcommand{\implBase}{(27.5,88.88888888888889)}%
\newcommand{\implSPump}{(27.5,44.44444444444444)}%
\newcommand{\implMPump}{(27.5,41.66666666666667)}%
\hspace{.73cm}
\begin{tikzpicture}[trim axis left, trim axis right]
\begin{axis}[
xlabel={Throughput (\si{\mega\sample\per\second})},
xlabel style={
yshift=2pt,
font=\footnotesize,
},
ylabel={DSP utilization (\si{\percent})},
ylabel style={
yshift=-8pt,
font=\footnotesize,
},
tick label style={
font=\footnotesize,
},
grid=major,
grid style={densely dotted},
legend entries={\emph{Base}, \emph{S-Pump}, \emph{M-Pump}},
width=.326\textwidth,
height=.326\textwidth,
legend style={
font=\footnotesize,
legend pos=\legendPos,
inner xsep=2pt,
inner ysep=.5pt,
rounded corners,
},
ymin=0,
ymax=100,
xmin=0,
set layers,
]
\pgfplotstableread[col sep=comma]{data/\fileName.csv}\datatable

\addplot[const plot, myblue, on layer=axis foreground, very thick, restrict x to domain=0:\memBound, forget plot] table [x=throughput, y=dsp_base] {\datatable};
\addplot[const plot, myblue, on layer=axis foreground, very thick, dashed, forget plot, restrict x to domain=\memBound:inf] table [x=throughput, y=dsp_base] {\datatable};
\addplot[only marks, mark options={fill=myblue, draw=myblue, scale=1.5}, forget plot] coordinates {\implBase};
\addplot[const plot, myorange, very thick, restrict x to domain=0:\degSBound, restrict x to domain=0:\memBound, forget plot] table [x=throughput, y=dsp_s-pump] {\datatable};
\addplot[const plot, myorange, very thick, dashed, forget plot, restrict x to domain=0:\degSBound, restrict x to domain=\memBound:inf] table [x=throughput, y=dsp_s-pump] {\datatable};
\addplot[only marks, mark options={fill=myorange, draw=myorange, scale=1.5}, forget plot] coordinates {\implSPump};
\addplot[const plot, mygreen, very thick, restrict x to domain=0:\degMBound, restrict x to domain=0:\memBound, forget plot] table [x=throughput, y=dsp_m-pump] {\datatable};
\addplot[const plot, mygreen, very thick, dashed, forget plot, restrict x to domain=0:\degMBound, restrict x to domain=\memBound:inf] table [x=throughput, y=dsp_m-pump] {\datatable};
\addplot[only marks, mark options={fill=mygreen, draw=mygreen, scale=1.5}, forget plot] coordinates {\implMPump};

\addplot[myblue, very thick] coordinates {(-1, -1)};
\addplot[myorange, very thick] coordinates {(-1, -1)};
\addplot[mygreen, very thick] coordinates {(-1, -1)};
\end{axis}
\end{tikzpicture}%
\label{fig:vms_tp_vs_dsp}%
}%
\hfill\phantom{}
\caption{
\Acfp*{DSP} allocated for a given throughput. The \emph{M-Pump} 
designs are optimized using the proposed task-level multi-pumping 
technique. The \emph{M-Pump} designs are Pareto-optimal compared to 
the \emph{Base} designs, whose \ac*{DSP} utilization is constant since they are
optimized by tuning the clock frequency only, and to the \emph{S-Pump} designs,
which are optimized for area by changing both the \ac*{II} and the global clock 
frequency of the tasks.
The dashed lines represent the theoretical throughputs achievable with the
allocated \acp*{DSP}, which are unreachable in practice due to memory bandwidth
limitations. The dots show the design points implemented in \acl*{HW}.
}
\label{fig:tp_vs_mult}
\end{figure*}
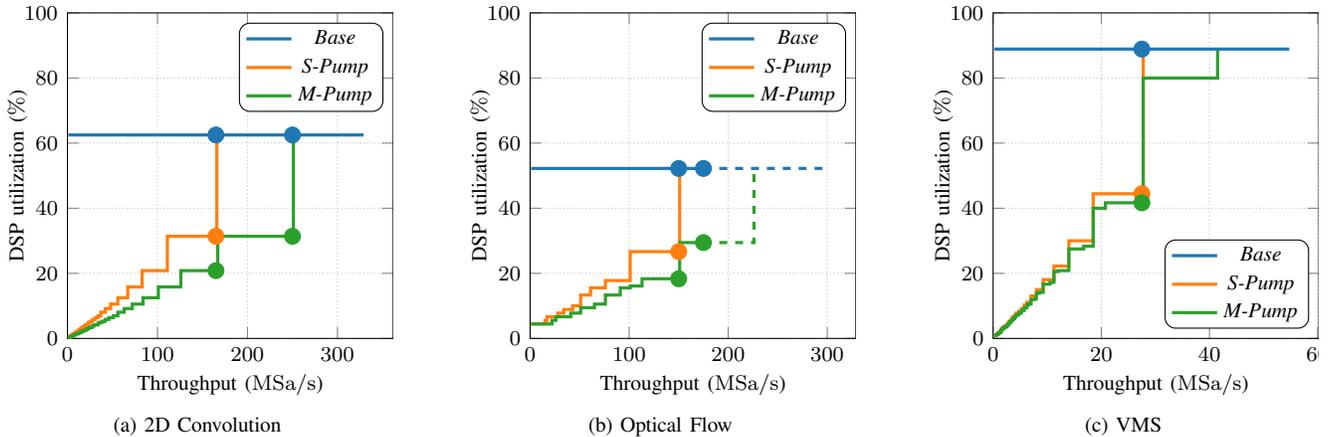

\Cref{fig:tp_vs_mult}
shows the tradeoffs between the \ac{DSP} utilization and the throughput
obtained by varying the base clock frequency within the range allowed by the
critical path of the designs.
The dashed lines represent computation throughputs that exceed 
the memory throughput. Thus, the effective throughputs are, in 
practice, clipped to the maximum non-dashed value, corresponding 
to the maximum memory throughput.

\begin{table*}
\footnotesize
\caption{\Acl*{PPA} of the benchmarks targeting a Zynq UltraScale+ \ac*{SoC}}
\label{tab:results}
\centering
\begin{tabular}{c|S[table-format=3.0]|c|cccccS[table-format=2.0]S[table-format=2.0]ccc}
\toprule
&&&&&& \multicolumn{2}{c}{\textbf{\acs*{LUT}}} &&&
\multicolumn{2}{c}{\textbf{Power}} & \\
& {\multirow{-2}*{\textbf{Throughput}}} &
&
\multirow{-2}*{\textbf{\shortstack{Base\\clock}}} &
\multirow{-2}*{\textbf{\shortstack{Pump\\factors}}} &
\multirow{-2}*{\textbf{\acs*{DSP}}} & \textbf{Logic} & \textbf{Memory} &
\multicolumn{1}{c}{\multirow{-2}*{\textbf{\acs*{FF}}}} & \multicolumn{1}{c}{\multirow{-2}*{\textbf{\acs*{BRAM}}}} &
\textbf{Static} & \textbf{Dynamic} &
\multirow{-2}*{\textbf{\shortstack{Clock\\routing}}} \\
\multirow{-3}*{\textbf{Design}} & {(\si{\mega\sample\per\second})} &
\multirow{-3}*{\textbf{Implem.}} &
(\si{\mega\hertz}) & (\si{\one}) & (\si{\percent}) & (\si{\percent}) & (\si{\percent}) &
\multicolumn{1}{c}{(\si{\percent})} & \multicolumn{1}{c}{(\si{\percent})} & (\si{\watt}) & (\si{\watt}) & (\si{\percent}) \\
\midrule
&& Base & 165 & -- & 64 & 14 & 12 & 9 & 4 & 0.3 & 1.8 & 1.0 \\
& 165 & S-Pump & 330 & 2 & 33 & 15 & 12 & 18 & 4 & 0.3 & 2.3 & 1.0 \\
&& M-Pump & 165 & 3 & \textbf{23} & 14 & 12 & 18 & 4 & 0.3 & 2.2 & 2.4 \\
\cmidrule{2-13}
&& Base & 250 & -- & 64 & 13 & 12 & 10 & 4 & 0.3 & 2.0 & 1.0 \\
\multirow{-5}*{\shortstack{2D Convolution\\\cite{xilinx_conv}}} & {\multirow{-2}*{250}} & M-Pump & 250 & 2 & \textbf{33} & 15 & 12 & 19 & 4 & 0.3 & 2.8 & 2.4 \\
\midrule
&& Base & 150 & -- & 55 & 36 & 65 & 23 & 20 & 0.3 & 2.5 & 1.0 \\
& 150 & S-Pump & 300 & 2 & 29 & 37 & 65 & 26 & 20 & 0.3 & 3.2 & 1.0 \\
&& M-Pump & 150 & 2, 3 & \textbf{21} & 37 & 64 & 27 & 20 & 0.3 & 3.0 & 3.8 \\
\cmidrule{2-13}
&& Base & 175 & -- & 55 & 36 & 65 & 23 & 20 & 0.3 & 2.5 & 1.0 \\
\multirow{-5}*{\shortstack{Optical Flow\\\cite{rosetta}}} & {\multirow{-2}*{175}} & M-Pump & 175 & 2 & \textbf{33} & 38 & 65 & 27 & 20 & 0.3 & 2.9 & 2.4 \\
\midrule
&& Base & 110 & -- & 89 & 32 & 23 & 31 & 67 & 0.3 & 2.0 & 1.0 \\
& 28 & S-Pump & 220 & 2 & 44 & 30 & 23 & 31 & 67 & 0.3 & 2.4 & 1.0 \\
\multirow{-3}*{\shortstack{VMS\\\cite{vms}}} && M-Pump & 110 & 2, 3 & \textbf{42} & 32 & 23 & 37 & 67 & 0.3 & 2.9 & 3.8 \\
\bottomrule
\end{tabular}
\end{table*}

The number of \acp{DSP} used by \emph{Base} designs is 
independent of the clock frequency. The plots of the \emph{Pump} 
designs are characterized by a step shape, whose discontinuities 
correspond to the \acp{II} changes, which only assume integer 
values. The \emph{Pump} solutions provide different tradeoffs in 
the throughput versus \ac{DSP} space, thanks to the tuning of the 
pipelines' \ac{II}. The additional degree of freedom of the 
\emph{M-Pump} implementations (i.e., the task clock frequency) 
makes them always Pareto optimal.

Both \emph{M-Pump} and \emph{S-Pump} designs degenerate to 
\emph{Base} designs (i.e.,  all the pumping factors to one and no 
resource  savings) at the highest throughputs since they need the 
lowest \acp{II} to reach the best performance. Note that the 
\emph{M-Pump} designs consistently degenerate to \emph{Base} at 
throughputs higher than \emph{S-Pump} since the multiple clock 
domains let the multi-pumped tasks run at the maximum frequency 
their local critical path allows. Therefore, the \emph{M-Pump} 
designs achieve up to \SI{52}{\percent} higher throughput than 
\emph{S-Pump} with the same \acp{DSP} in the 2D Convolution test 
case. Moreover, with the Optical Flow benchmark, the 
\emph{M-Pump} reaches the maximum effective throughput using 
\SI{40}{\percent} fewer \acp{DSP} than the \emph{Base}.

\Cref{tab:results} reports the post-implementation \ac{PPA} data 
for the design points marked with the dots  in 
\cref{fig:tp_vs_mult}. We select those points since their 
throughputs are the upper extremes of the last steps of 
\emph{M-Pump} and \emph{S-Pump} within the memory bound. 

Comparing the \emph{M-Pump} designs with the \emph{Base} ones, 
the consistent \ac{DSP} saving (\SI{54}{\percent} on average) 
implies power and \acp{FF} overheads. The additional power 
(\SI{24}{\percent} on average) is because the multi-pumped tasks 
are characterized by greater switching activity due to higher 
resource reuse and clock frequencies. The additional \acp{FF} 
(\SI{33}{\percent} on average) are inserted by the \ac{HLS} tool 
in the multi-pumped tasks to build deeper pipelines and reach 
higher clock frequencies.

As expected \cite{fifo_gen_pru}, the \ac{PPA} overhead for 
\ac{CDC} in \emph{M-Pump} is negligible. The overhead for routing 
multiple clocks is also marginal, as each additional clock domain 
allocates only \SI{1.4}{\percent} of the available clock routing 
resources.

In general, the \emph{M-Pump} solutions Pareto dominate the 
\emph{S-Pump} ones. In fact, at the same throughput, they 
allocate fewer \acp{DSP}, similar \acp{LUT} and \acp{FF}, and 
consume less power. This is because the \emph{M-Pump} designs 
take advantage of the multiple clock domains to increase the 
clock frequency of the multi-pumped tasks only, thus reaching 
higher multi-pumping factors and avoiding power and \ac{FF} 
overheads in the non-multi-pumped tasks. The \ac{VMS} test case 
is the only exception because only a small fraction of its logic 
runs at the base clock frequency, while the rest is double or 
triple-pumped; thus, the lower-frequency tasks are not enough to 
balance the power and \ac{FF} overhead for the multi-pumped 
tasks.

\section{Conclusion}
\label{sec:conclusion}

We propose a task-level multi-pumping technique for saving 
\acl{HW} resources while maintaining the original throughput for 
\ac{HLS} dataflow designs for \acp{FPGA}.

Given a \acl{SOTA} single-clock \ac{DFG}, our approach first 
generalizes it to a multi-clock \ac{DFG}. Secondly, it tunes the 
tasks' high-level parameters (i.e., clock frequency and pipeline 
\ac{II}) to multi-pump their \aclp{FU}. The overhead for 
generalization is negligible, thanks to the \acp{DFG} structure, 
which consists of independent blocks communicating via 
\acp{FIFO}, allowing for safe \ac{CDC}, and modern \ac{FPGA} 
clock architectures, which seamlessly handle multiple clock 
domains even if current \ac{HLS} tools do not exploit them.

The experimental results reported in \cref{sec:bench} prove that 
our method opens a new Pareto front in the performance versus 
\acp{DSP} space, saving up to \SI{40}{\percent} of resources at 
maximum throughput. Moreover, our method does not require any 
manual architecture changes from the designer, since it acts only 
on the high-level parameters of the tasks and uses the \ac{HLS} 
binding algorithm to automatically generate the resource sharing 
logic. Finally, the generalization to multi-clock \acp{DFG} 
simply requires replacing single-clock with multi-clock 
\acp{FIFO}. Therefore, our technique is well suited for a fully 
automated \ac{HLS} optimization pass, which will be the subject 
of future work.

\section*{Acknowledgment}

This work was partially supported by the Key Digital Technologies 
Joint Undertaking under the REBECCA Project with grant agreement 
number 101097224, receiving support from the European Union, 
Greece, Germany, Netherlands, Spain, Italy, Sweden, Turkey, 
Lithuania, and Switzerland.



\IEEEtriggeratref{23}
\printbibliography

\end{document}